\documentclass{emulateapj}
\usepackage{apjfonts}

\usepackage{lscape}
\usepackage{graphicx}
\usepackage{natbib}
\usepackage{longtable}

\newcommand{\hbeta}{H{$\beta$}}

\newcommand{\CIV}{C{\sevenrm IV}}

\def\MgII{Mg\,{\sc ii}}

 \font\sevenrm=cmr7 scaled 1000

\begin{document}

\title{Astrometric Reverberation Mapping}

\shorttitle{Astrometric RM}

\slugcomment{ApJ in press}

\shortauthors{SHEN}
\author{Yue Shen\\
Harvard-Smithsonian Center for Astrophysics, 60 Garden
Street, MS-51, Cambridge, MA 02138, USA}

\begin{abstract}

Spatially extended emission regions of Active Galactic Nuclei (AGN) respond
to continuum variations, if such emission regions are powered by energy
reprocessing of the continuum. The response from different parts of the
reverberating region arrives at different times lagging behind the continuum
variation. The lags can be used to map the geometry and kinematics of the
emission region (i.e., reverberation mapping, RM). If the extended emission
region is not spherically symmetric in configuration and velocity space,
reverberation may produce astrometric offsets in the emission region
photocenter as a function of time delay and velocity, detectable with future
$\mu$as to tens of $\mu$as astrometry. Such astrometric responses provide
independent constraints on the geometric and kinematic structure of the
extended emission region, complementary to traditional reverberation mapping.
In addition, astrometric RM is more sensitive to infer the inclination of a
flattened geometry and the rotation angle of the extended emission region.

\end{abstract}


\section{Introduction}\label{sec:intro}

Reverberation mapping (RM) is a powerful tool to probe the structure of the
broad emission line region (BLR) in AGNs and quasars
\citep[e.g.,][]{Blandford_McKee_1982,Peterson_1993} without the need to
resolve the BLR. The basic ideas are that the BLR is powered by
photoionization by the (ionizing) continuum from the accreting black hole and
that it responds to the variations of the continuum in a light crossing time.
The response from different parts of the BLR will arrive at different time
delays with respect to the continuum variation. Thus by mapping a
two-dimensional broad line response function in the time delay and velocity
plane \citep[velocity-delay maps,
e.g.,][]{Blandford_McKee_1982,Horne_etal_2004} one can in principle recover
the geometry and kinematics of the BLR. Over the past several decades, RM has
proven to be a practical technique in studying the structure of BLRs.
Although accurate velocity-time delay mapping is still lacking, RM studies
have successfully measured average BLR sizes for several dozens of AGNs and
quasars
\citep[e.g.,][]{Kaspi_etal_2000,Kaspi_etal_2007,Peterson_etal_2004,Bentz_etal_2009a,Bentz_etal_2009b},
and in a few cases, crude velocity-delay maps
\citep[e.g.,][]{Denney_etal_2009a,Bentz_etal_2010}.

RM studies have shown that the typical BLR size $R$ scales approximately with
the AGN luminosity $L^{0.5}$. In the latest version of the $R-L$ relation
\citep[e.g.,][]{Bentz_etal_2009a}, $\log (R/{\rm light\,
days})=-21.3+0.519\log(\lambda L_{\lambda}(5100\textrm{\AA})/{\rm
erg\,s^{-1}})$. Thus for typical quasar luminosities ($\lambda
L_{\lambda}(5100\textrm{\AA})=10^{45}\,{\rm erg\,s^{-1}}$, or bolometric
luminosity $L_{\rm bol}\approx 10^{46}\,{\rm erg\,s^{-1}}$) at $z\sim 0.5$,
the BLR size is $\sim 0.1\,{\rm pc}$ ($\sim 15\,\mu$as). This scale is 3
orders of magnitude smaller than the diffraction limit of 10m class optical
telescopes (${\cal \theta}\sim$ tens of mas). Optical/near-IR interferometry
with $\mu$as resolution has yet to come. Thus reverberation mapping will
remain one of the few practical methods to measure the BLR size, along with
microlensing in gravitationally lensed quasars
\citep[e.g.,][]{Morgan_etal_2010, Sluse_etal_2011}.

However, measuring the source photocenter positions can achieve a factor of
$\sim 1/\sqrt{N_{\rm photon}}$ enhancement in precision compared with the
image resolution \citep[e.g.,][]{Lindegren_1978,Bailey_98a}, where $N_{\rm
photon}$ is the number of photons received in a bandpass. This means for
$10^6$ photons, the achievable astrometry precision is 3 orders of magnitude
smaller than the image resolution. One application of this idea is {\em
spectroastrometry} \citep[e.g.,][]{Bailey_98a}, where differential
astrometric positions as a function of wavelength (velocity) can be used to
probe otherwise unresolved sources
\citep[e.g.,][]{Bailey_98b,Gnerucci_etal_2011}.

When the astrometric precision approaches the angular BLR size, it becomes
possible to study the BLR structure with astrometric signatures. A simple
application would be to use spectroastrometry to place constraints on the BLR
structure\footnote{For instance, in certain geometries of the BLR, the blue
and red parts of the broad line emission may have offset photocenters. The
detection of such photocenter offsets can put constraints on the BLR size and
kinematics. }. A more radical possibility, however, is to detect and model
the wobble in the broad line emission photocenter due to reverberation to the
continuum variations, since the different arrival times of the BLR response
will cause shifts in the observed BLR photocenter.

In this work we investigate the feasibility of combing the traditional
intensity RM with astrometric information. Unlike spectroastrometry, the
reverberation of the BLR will directly induce shifts in the photocenter
position of broad line emission as a function of time delay and velocity. As
we will show below, the pattern of photocenter shifts is determined by the
geometry and kinematics of the BLR, thus providing independent constraints on
the BLR structure, complementary to traditional intensity RM. Although our
main focus is on the BLR, this method can be readily applied to the
reverberation of the dust torus of AGNs \citep[e.g.,][]{Suganuma_etal_2006}
or other reverberation systems, where the required astrometric precision
might be less stringent. We describe the basic formalism in \S\ref{sec:form}
and demonstrate this method in \S\ref{sec:model} and \S\ref{sec:result} with
simple models for the BLR. We briefly discuss the practical issues with this
method in \S\ref{sec:disc}, with an overview on the perspectives of achieving
the required astrometric precision in the near future in
\S\ref{sec:disc_astro}, and conclude in \S\ref{sec:con}.

\section{Basic Formalism}\label{sec:form}

Let us set up a 3-dimensional cartesian coordinate system (in the observer
frame) centered on the continuum source (e.g., the inner edge of the
accretion disk, which we treat as a point source), and let the observer be at
$z=+\infty$. The time delay of response from each point $\mathbf{r}\equiv
(x,y,z)$ in the BLR is then simply:
\begin{equation}\label{eqn:time_delay}
\tau(\mathbf{r}) = r-z=\sqrt{x^2+y^2+z^2}-z=r(1-\cos\theta)\ ,
\end{equation}
where we use units that normalize the speed of light to $c=1$, $r$ is the
distance to the central source, and $\theta$ is the angle between vector
$\mathbf{r}$ and $+z$ axis. Obviously the iso-delay surface (i.e., the
surface with constant time delay) is a paraboloid for an observer located at
infinity. Points on the $+z$ axis, i.e., on the near side along the
line-of-sight (hereafter los), have zero time delay with respect to the
continuum variations.

Assuming that the observed continuum increases at time $t=0$ by a constant
amount $f_c$ for a certain amount of time $\Delta t$, and goes back to normal
at $t=\Delta t$, i.e., the change in continuum luminosity is
\begin{eqnarray}\label{eqn:conti_vary}
\delta_c(t)  & = &
 \left\{
\begin{array}{ll}
0,\quad  & t<0 \\
f_c,\quad  & 0<t<\Delta t \\
0,\quad  & t>\Delta t\ .
\end{array} \right.
\end{eqnarray}
The response in the broad line intensity is then a function of time $t$ and
los velocity $v$
\begin{equation}\label{eqn:rm_amp}
\delta_{\rm em}(t,v) = \int j(\mathbf{r})g(v,\mathbf{r},\mathbf{w})\delta_{c}\left(t-\tau(\mathbf{r})\right)d\mathbf{r}d\mathbf{w}\ ,
\end{equation}
where $j(\mathbf{r})$ is the (assumed isotropic) responding volume emissivity
of the emission region as a function of position, and
\begin{equation}
g(v,\mathbf{r},\mathbf{w})=f(\mathbf{r},\mathbf{w})\delta(v-\mathbf{w}\cdot \mathbf{n})\
\end{equation}
where $(\mathbf{r,w})$ are the 3D coordinate and velocity vectors,
$f(\mathbf{r},\mathbf{w})$ is the normalized (e.g., $\int fd\mathbf{w}=1$) 3D
velocity distribution at point $\mathbf{r}$, determined by the kinematic
structure of the BLR, and $\mathbf{n}\equiv +\hat{z}$ denotes the unit vector
of the los. Note that Eqn.\ (\ref{eqn:rm_amp}) is equivalent to the
traditional RM equation invoking the transfer function \citep[i.e., the line
intensity response for a $\delta$-function continuum
flare;][]{Blandford_McKee_1982,Peterson_1993}. However, for simplicity we
shall avoid the usage of transfer function and work directly in the time
domain in terms of the observed line intensity and photocenter positon for
different velocity channels \citep[see, e.g.,][]{Pancoast_etal_2011}.

If we further assume that the BLR is transparent to its own emission, i.e.,
neglecting absorption and scattering (as generally assumed in RM), then the
projected photocenter position of the {\em responding} emission with los
velocity $v$ as observed by the distance observer is
\begin{eqnarray}\label{eqn:rm_astro}
x_0(t,v)&=&\displaystyle\frac{\int xj(\mathbf{r})g(v,\mathbf{r},\mathbf{w})\delta_{c}\left(t-\tau(\mathbf{r})\right)d\mathbf{r}d\mathbf{w}}
{\int j(\mathbf{r})g(v,\mathbf{r},\mathbf{w})\delta_{c}\left(t-\tau(\mathbf{r})\right)d\mathbf{r}d\mathbf{w}}\nonumber\\
y_0(t,v)&=&\displaystyle\frac{\int yj(\mathbf{r})g(v,\mathbf{r},\mathbf{w})\delta_{c}\left(t-\tau(\mathbf{r})\right)d\mathbf{r}d\mathbf{w}}
{\int j(\mathbf{r})g(v,\mathbf{r},\mathbf{w})\delta_{c}\left(t-\tau(\mathbf{r})\right)d\mathbf{r}d\mathbf{w}}\ .
\end{eqnarray}

The projected photocenter position integrated over all los velocities is:
\begin{eqnarray}\label{eqn:rm_astro_all}
x_0(t)=\displaystyle\frac{\int xj(\mathbf{r})\delta_{c}\left(t-\tau(\mathbf{r})\right)d\mathbf{r}}
{\int j(\mathbf{r})\delta_{c}\left(t-\tau(\mathbf{r})\right)d\mathbf{r}}\ ,
y_0(t)=\displaystyle\frac{\int yj(\mathbf{r})\delta_{c}\left(t-\tau(\mathbf{r})\right)d\mathbf{r}}
{\int j(\mathbf{r})\delta_{c}\left(t-\tau(\mathbf{r})\right)d\mathbf{r}}\ .\nonumber \\
\end{eqnarray}

In Eqn.\ (\ref{eqn:conti_vary}), if $c\Delta t$ is much smaller compared with
the typical BLR size $R$, then the continuum light curve can be treated as a
(square wave) $\delta-$function. In this case the near side of the BLR
completes reverberation (i.e., back to the steady level) before the far side
reverberation reaches the observer. On the other hand, if $c\Delta t\gg R$,
then the continuum light curve is a step function, and the near side of the
BLR will still be in the high state when the far side reverberation reaches
the observer. In the following discussion, we will consider the two limiting
cases where $c\Delta t\ll R$ (a continuum flare) and $c\Delta t\gg R$ (a
continuum jump).

\section{Simple Geometric and Kinematic Models for the BLR}\label{sec:model}

\begin{figure*}
 \centering
 \includegraphics[width=0.32\textwidth]{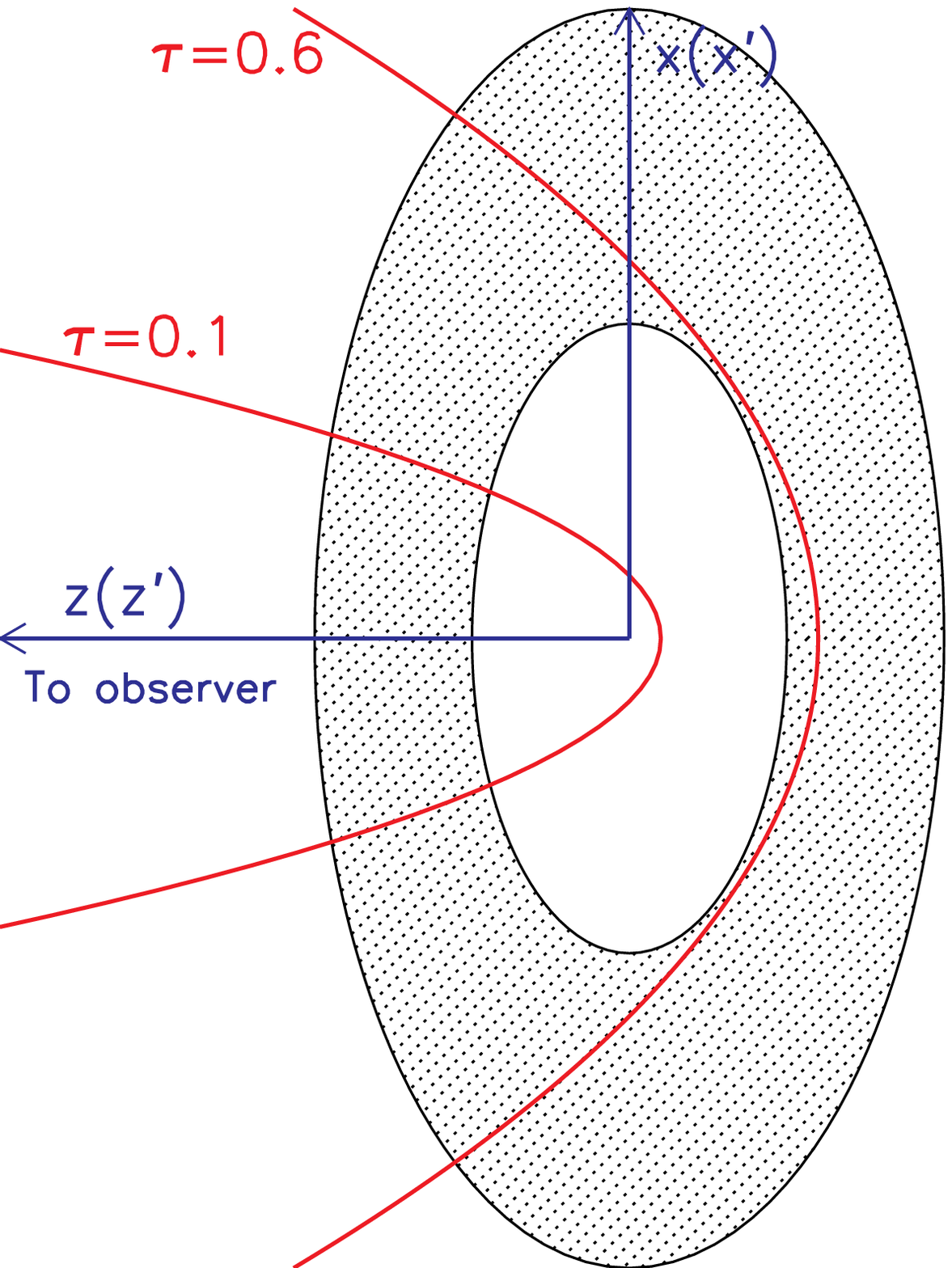}\vspace{2mm}
 \includegraphics[width=0.32\textwidth]{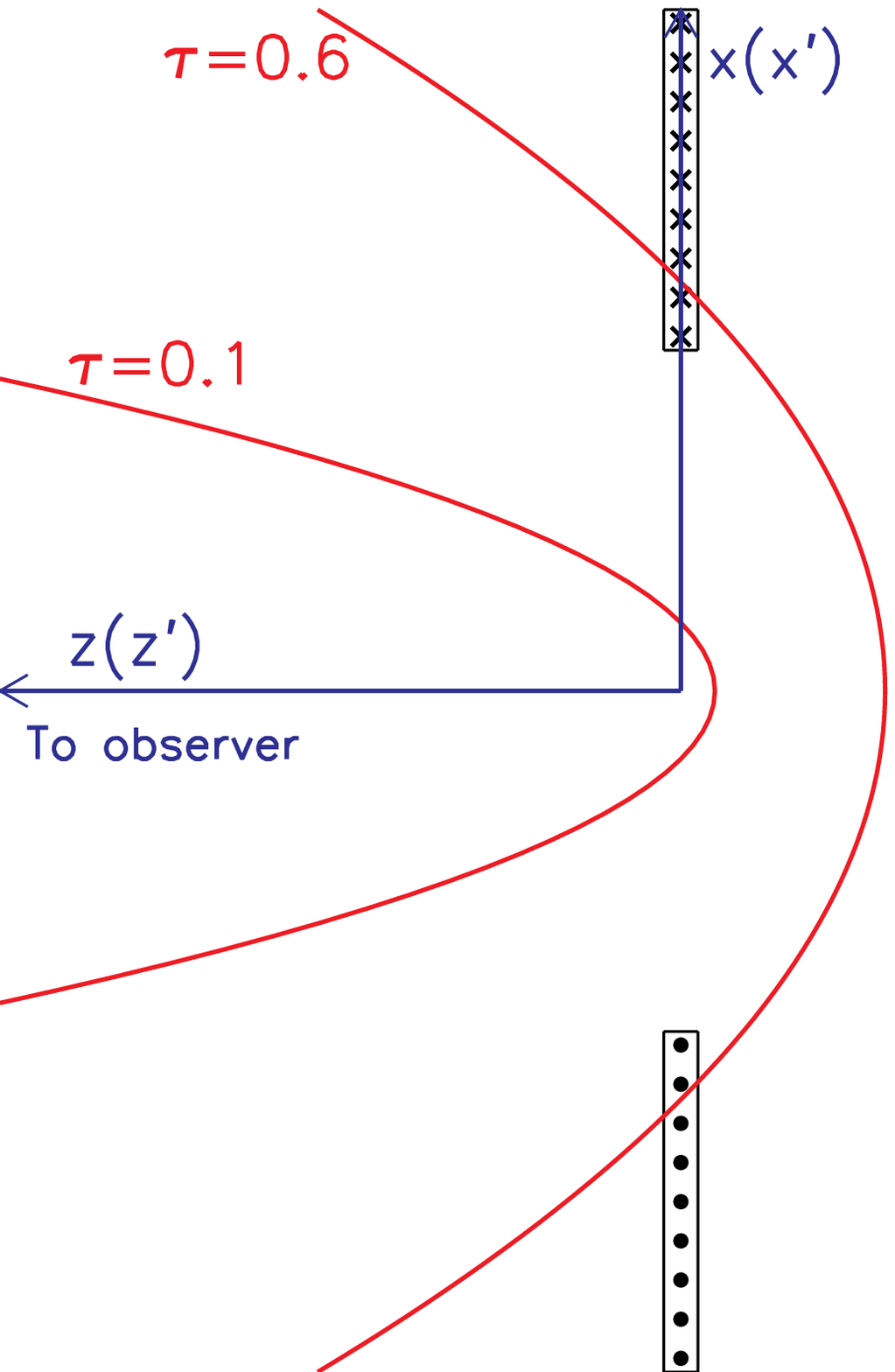}\vspace{2mm}
 \includegraphics[width=0.32\textwidth]{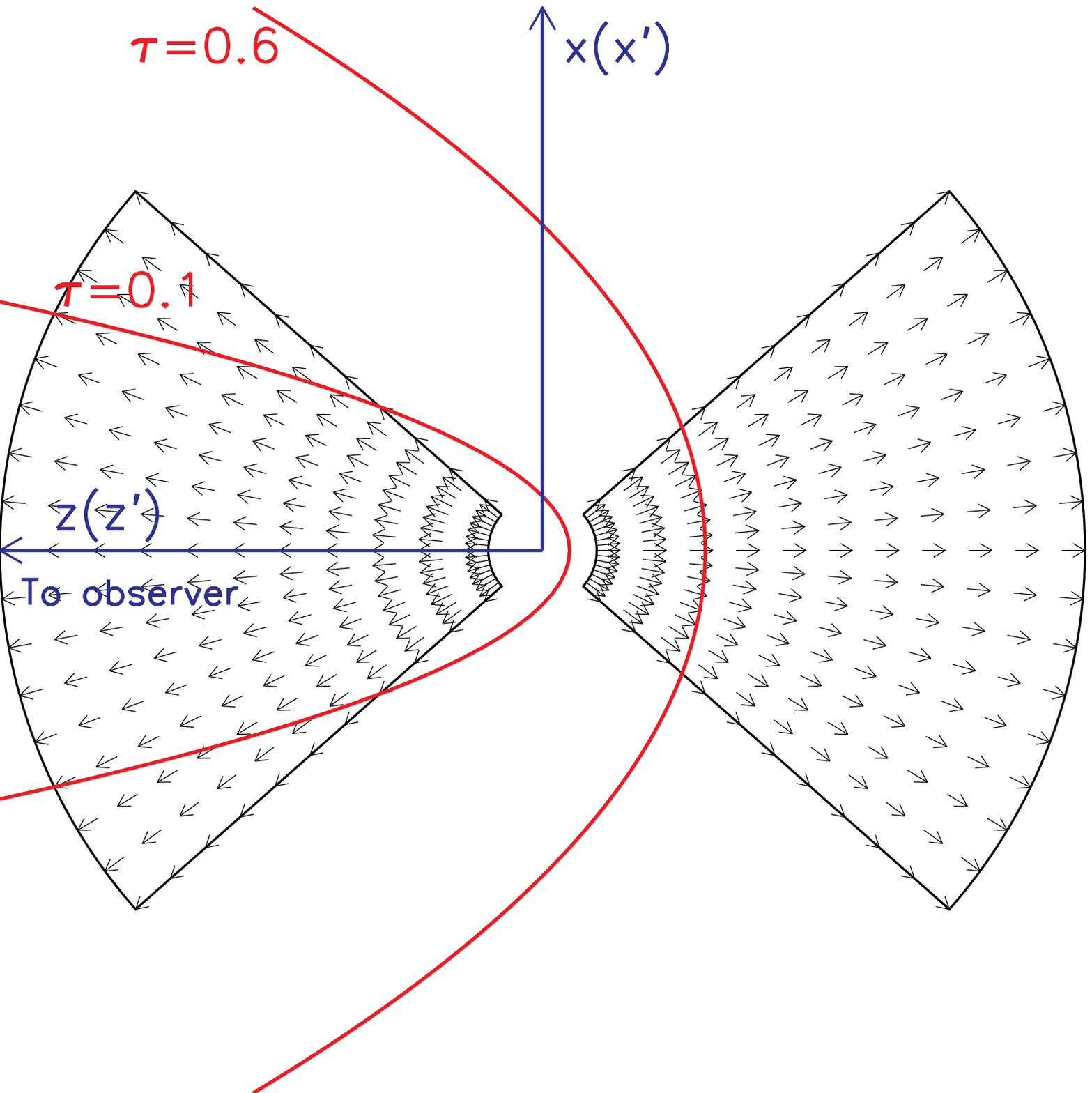}\vspace{2mm}
 \includegraphics[width=0.32\textwidth]{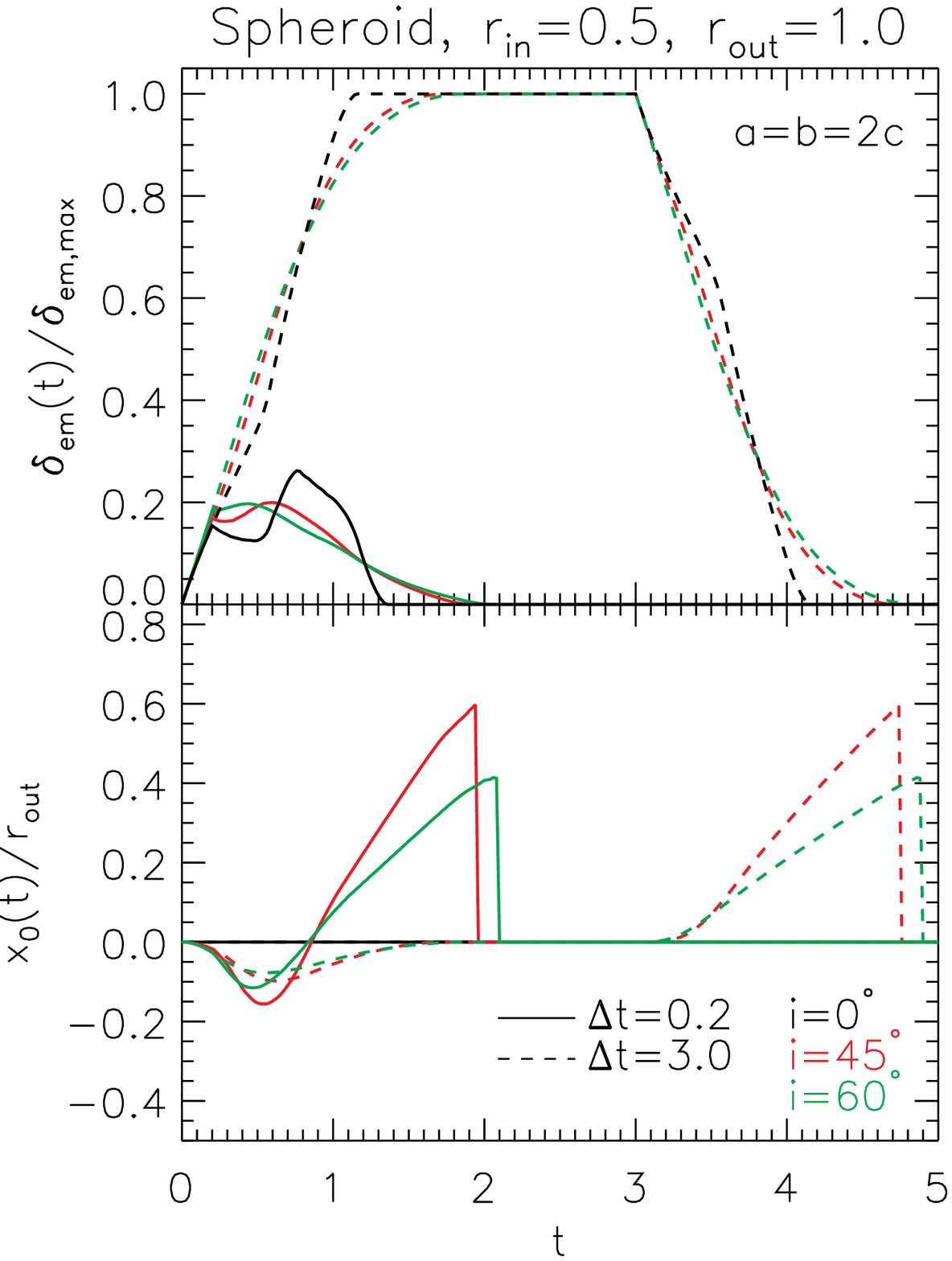}
 \includegraphics[width=0.32\textwidth]{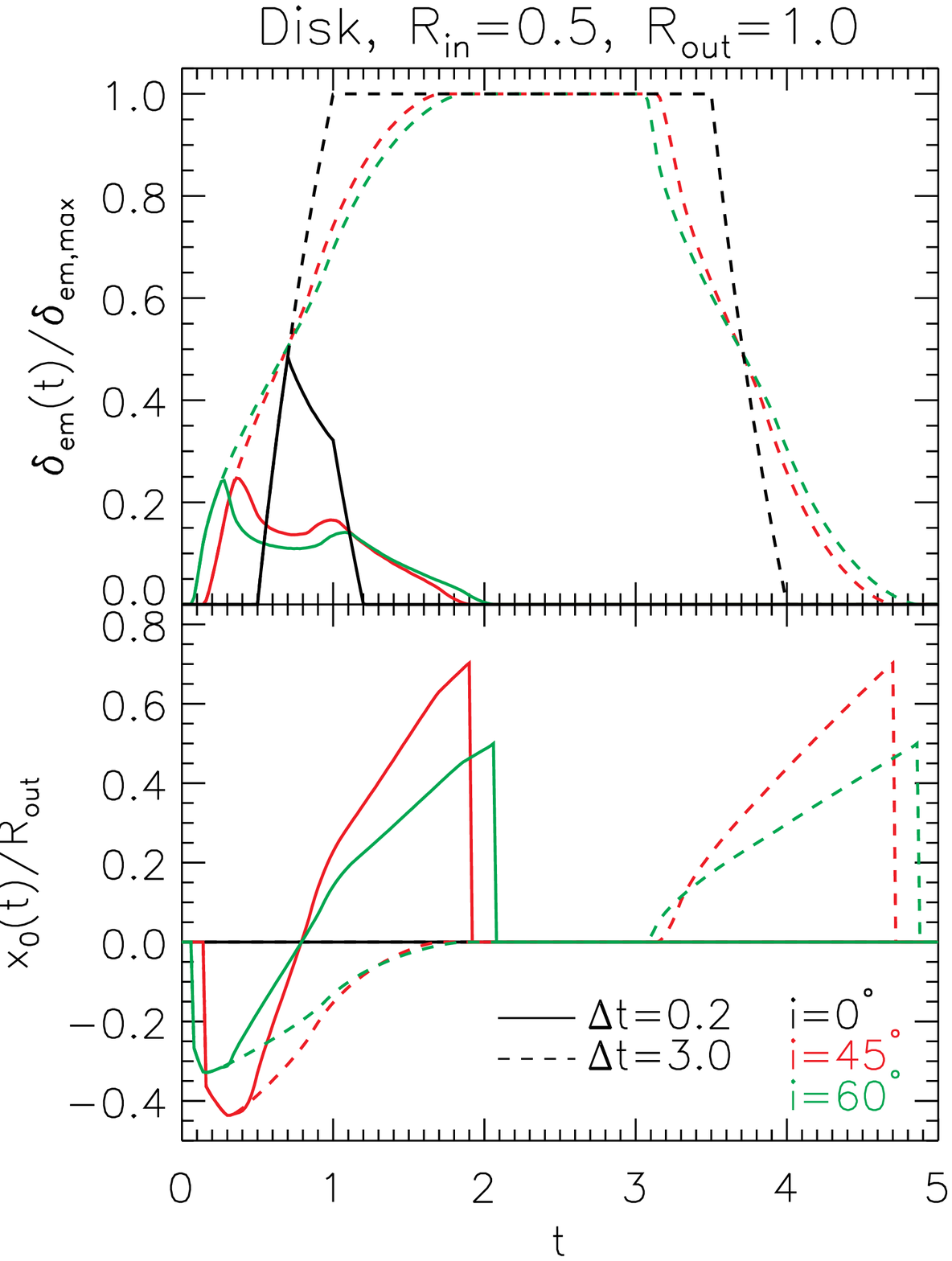}
 \includegraphics[width=0.32\textwidth]{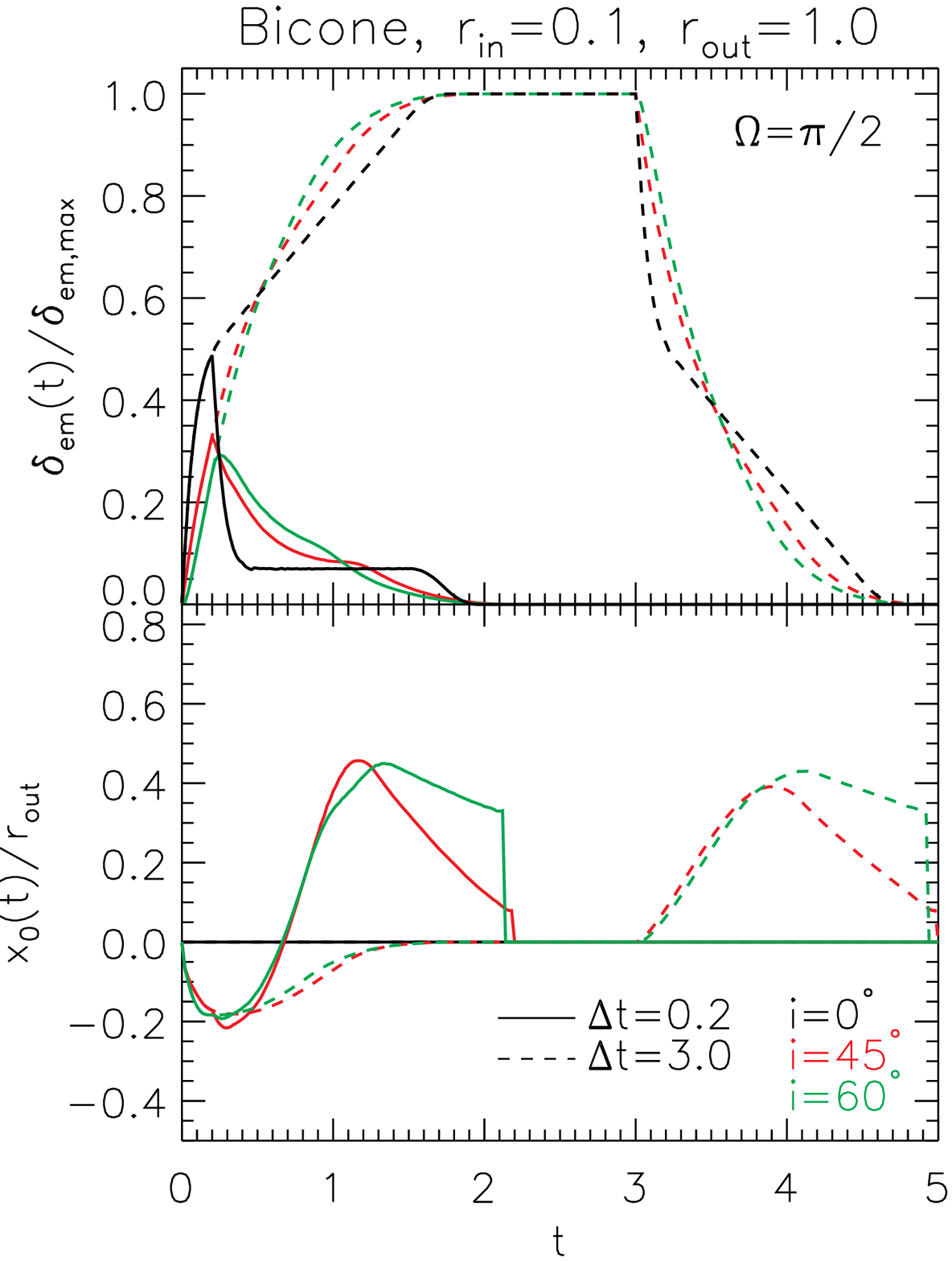}
    \caption{Illustrative examples of BLR geometry and kinematics for the three
    non-spherical models described in \S\ref{sec:model}, and for the velocity-integrated
    reverberation amplitude $\delta_{\rm em}(t)$ and photocenter position $x_0(t)$. The top panels show
    the cross section (regions filled with different patterns) in the $xz$ plane for three BLR geometries
    when the BLR intrinsic coordinate system $x^\prime y^\prime z^\prime$
    is aligned with the observer coordinate system $xyz$. The red curves are typical parabolic iso-delay
    curves with constant $\tau$ (in the same units as in Eqn.\ \ref{eqn:time_delay}). The bottom panels
    show the corresponding reverberation mapping signals. {\em
    Left:} an oblate spheroid model with $a=b=2c=1$, $r_{\rm in}=0.5$ and $r_{\rm
    out}=1$. {\em Middle:} a truncated disk model with $R_{\rm in}=0.5$ and
    $R_{\rm out}=1$. {\em Right:} a biconical outflow with $r_{\rm in}=0.1$,
    $r_{\rm out}=1.0$, and an opening solid angle $\Omega=\pi/2$. In all
    examples, the intrinsic coordinates of the BLR ($x^\prime y^\prime z^\prime$)
    are rotated about the $y$-axis in the observer frame by an inclination angle $i$,
    such that the projected BLR in the $xy$ plane has planar symmetry about the
    $x$-axis (hence $y_0(t)\equiv 0$). For each model we compute for three inclination angles
    $i=0^\circ,45^\circ,60^\circ$, and two continuum variation timescales
    $\Delta t=0.2$ and $0.3$. In the $i=0^\circ$ cases, there is no
    photocenter offset as a function of time due to symmetry. We define the
    maximum-possible reverberation intensity as the one where the entire BLR
    is reverberating.
    }
    \label{fig:rm_examp}
\end{figure*}

\begin{figure}
 \centering
 \includegraphics[width=0.48\textwidth]{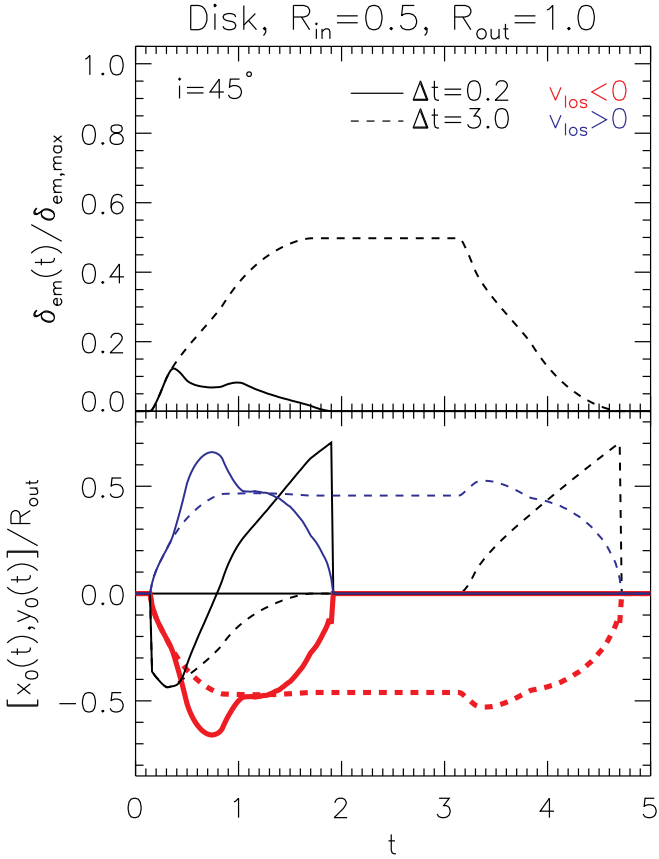}
    \caption{Velocity-resolved evolution of reverberation intensity and photocenter
    positions, for the same disk model as in Fig.\ \ref{fig:rm_examp} with a los angle
    $i=45^\circ$ and two continuum variation timescales $\Delta t=0.2,3$. Due to symmetry, the intensity $\delta_{\rm em}$ and $x$-positions
    of the photocenter have identical time evolution for the blueshifted and redshifted
    halves of the disk, which are denoted by the black lines. On the other hand, the
    $y$ photocenter positions have time evolutions denoted by the blue and (thicker) red lines for
    the blueshifted and redshifted halves, which are symmetric about the $x$-axis.
    }
    \label{fig:rm_vel_examp}
\end{figure}
\begin{figure}
 \centering
 \includegraphics[width=0.48\textwidth]{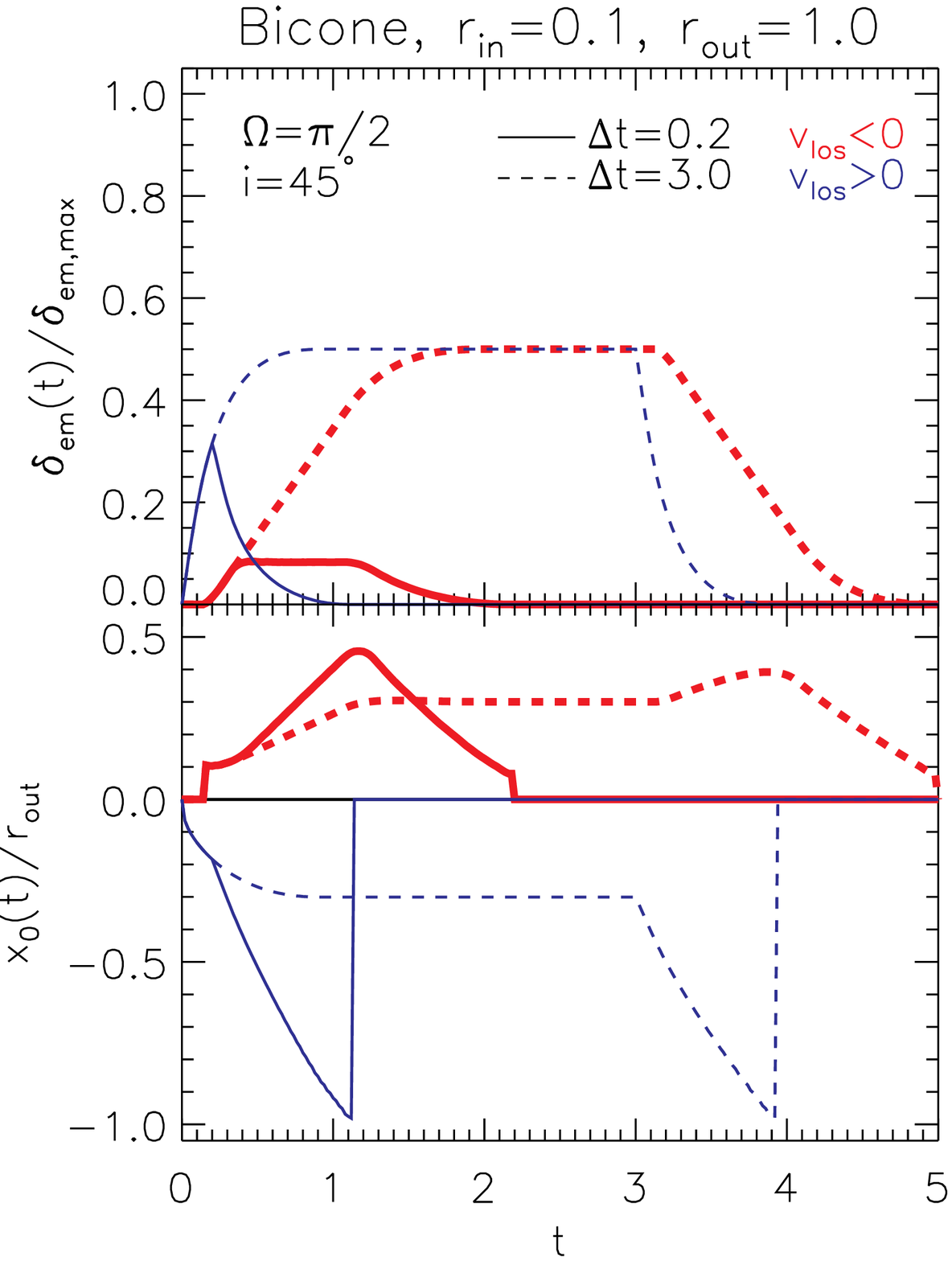}
 \caption{Velocity-resolved evolution of reverberation intensity and photocenter
    position $x_0(t)$, for the same bicone model as in Fig.\ \ref{fig:rm_examp} with a los angle
    $i=45^\circ$ and two continuum variation timescales $\Delta t=0.2,3$. The blueshifted
    cone (denoted by blue lines) and the redshifted cone (denoted by thicker red lines) have quite different time evolution in reverberation
    intensity $\delta_{\rm em}$ and photocenter position $x_0(t)$, reflecting the geometry
    and kinematics of the bicone BLR structure. The photocenter position $y_0(t)\equiv 0$
    due to symmetry in the projected $xy$ plane. }
 \label{fig:rm_vel_bicone_examp}
\end{figure}

To demonstrate the astrometric signals from RM, we use simple geometric and
kinematic models for the BLR, and study the reverberation process. But our
approach below is applicable to general cases.

There are few observational constraints on the structure of the BLR.
Spherical BLRs will not produce any photocenter offset. However, there is
evidence that the BLR motion may be dominated by rotation, and the BLR may
have a flattened geometry. Such evidence comes from the dependence of broad
line width on orientation angle inferred from radio properties
\citep[e.g.,][]{Wills_Browne_1986,Jarvis_McLure_2006}, or from modeling the
broad line profiles \citep[e.g.,][]{Kollatschny_Zetzl_2011}. In the extreme
case, the broad lines in $\lesssim 10\%$ AGNs show a characteristic
double-peaked profile, commonly interpreted as arising from a disk geometry
with Keplerian rotation \citep[e.g.,][]{Eracleous_Halpern_1994}. On the other
hand, radial motion of the BLR has been inferred from velocity-resolved
reverberation mapping in several AGNs \citep[e.g.][]{Denney_etal_2009a}, and
it has been suggested that the high-ionization broad line (such as \CIV) may
have a component originating from a disk wind
\citep[e.g.,][]{Murray_etal_1995,Proga_etal_2000}. It is possible that virial
motion and radial motion coexist in the BLR.

Motivated by these observations, we consider the following simple classes of
geometry and kinematics for the BLR\footnote{Note that in this paper we do
not intend to construct fully self-consistent models for the BLR, nor do we
intend to reproduce the observed emission line profile with our BLR models.
Rather, we use these idealized models to demonstrate the expected signals
from reverberation mapping. More realistic BLR models can be easily
implemented in our framework. }. In all cases we assume constant density of
the BLR, and we assume uniform reprocessing coefficient $\epsilon$ across the
entire BLR. Therefore the volume emissivity $j(\mathbf{r})\equiv \epsilon
L_c/(4\pi r^2)$ is proportional to $1/r^2$, i.e., spherical shells with the
same thickness but different radii will have the same total reverberation
intensity. Again, we emphasize that our approach is not limited to these
assumptions. We establish an intrinsic coordinate system $(x^\prime y^\prime
z^\prime)$ anchored to the BLR, which is rotated from the observer coordinate
system $(x,y,z)$ by Euler angles $(\alpha, \beta, \gamma)$.

\begin{enumerate}
\item A triaxial ellipsoid configuration distributed within a shell:
    $r_{\rm in}^2<\displaystyle\frac{x^{\prime 2}}{a^2}+\frac{y^{\prime
    2}}{b^2}+\frac{z^{\prime 2}}{c^2}<r_{\rm out}^2$. We further consider
    a sub-class of this configuration, an oblate spheroid where $a=b>c$,
    which probably more represents a flattened BLR geometry. The velocity
    field is assumed to be Maxwellian, with 1D dispersion equal to the
    virial velocity at each radius. Spherical shell geometries are also
    simplified versions of this model.

\item A truncated circular razor-thin disk ($R_{\rm in}<R<R_{\rm out}$)
    in the $x^\prime y^\prime$ plane, with an inclination angle $i$ from
    the los ($z$-axis in the observer frame). We assume a Keplerian
    velocity field for the disk and the disk rotation is
    counter-clockwise viewed from $z^\prime =+\infty$.

\item A biconical configuration, which is rotationally symmetric about
    the $z^\prime$-axis with an opening solid angle $\Omega$, and a
    radial extent from $r_{\rm in}$ to $r_{\rm out}$. The bicone has an
    inclination angle $i$ from the los ($z$-axis in the observer frame).
    We assume a biconical radial outflowing velocity structure. The
    detailed outflow velocity structure is not important in this work, as
    we will only consider the blueshifted and redshifted halves of the
    broad line profile.

\end{enumerate}

To study reverberation mapping in emission line intensity and photocenter
offsets for arbitrary geometry and kinematics, we discretize the emission
region with a fine cartesian grid of cells $(x_i,y_i,z_i)$ and assign
velocities to each cell. At each time $t$, we determine the cells that are
reverberating, i.e., $\delta_c(t-\tau(x_i,y_i,z_i))\ne 0$, and compute the
reverberation quantities $\delta_{\rm em}$, $x_0$ and $y_0$ in Eqns.\
(\ref{eqn:rm_amp}) and (\ref{eqn:rm_astro}) with direct summation. This
discrete method can efficiently handle any geometry and kinematics to
sufficient numeric accuracy\footnote{In all the calculations shown below we
have performed convergence tests to make sure that the grid is fine enough to
capture the results to $<1\%$ relative accuracy.} compared with integrating
Eqns.\ (\ref{eqn:rm_amp}) and (\ref{eqn:rm_astro}). For simplicity we adopt
the units $c=G=M=1$, where $G$ is the gravitational constant and $M$ is the
central black hole mass.

\section{Results}\label{sec:result}

In the simplest case of a thin spherical shell (with radius $R$) model for
the emission line region, the time evolution of the reverberation intensity
has simple analytical forms \citep[e.g.,][]{Bahcall_etal_1972}, which we
verified with our numerical approach. However, due to the spherical geometry,
the photocenter does not show any offset as a function of time.

Fig.\ \ref{fig:rm_examp} shows several examples for the three non-spherical
models described in \S\ref{sec:model}, for the velocity-integrated case.
Since all three models have rotational symmetry, we need only focus on the
photocenter offset in the $x$-direction, i.e., the rotation of our model BLR
preserves symmetry about the $x$-axis in the projected $xy$ plane.
Specifically, the intrinsic coordinate system of the BLR $(x^\prime y^\prime
z^\prime)$ is rotated about the $y$-axis in the observer frame by inclination
angle $i$.

For each model we compute for three orientations $i=0,45^\circ,60^\circ$ and
two continuum variability durations $\Delta t=0.2$ and $3$. In all cases the
maximum BLR radius is set to be 1. It is clear from Fig.\ \ref{fig:rm_examp}
that similar information is encoded in the photocenter reverberation maps as
in the traditional intensity reverberation map. However, the photocenter
reverberation maps are more sensitive to the inclination of the flattened BLR
geometry than the intensity reverberation maps. The maximum photocenter
offset is on the same order as the flux-weighted average size of the BLR. We
will return to these points in \S\ref{sec:disc}.

In the examples shown in Fig.\ \ref{fig:rm_examp}, a generic feature of the
photocenter reverberation is the transition from the near-side photocenter to
the far-side photocenter. If the duration of the continuum variation $\Delta
t$ is short compared with the BLR size (e.g., a flash), this transition is
continuous. On the other hand, if the duration $\Delta t$ is long, then there
will be a period with no photocenter offset (i.e., the entire BLR is lit up).
The spheroid geometry and the disk geometry result in similar astrometric RM
signatures. This is expected because the spheroid model used is close to a
flattened disk geometry. In both cases we see a gradual rise followed by a
sharp cutoff around the maximum photocenter offset after the continuum
returns to the base value. However, in the bicone geometry, the sharp cutoff
occurs later past the peak photocenter offset. This difference is caused by
the different edge geometry in these models -- in the disk and spheroid
cases, fewer and fewer materials are located at the far side of the emission
region, which is opposite in the bicone case.

Another point worth noting is that for inclined BLRs, the total line
intensity reverberation $\delta_{\rm em}(t)$ looks similar in different
models, thus does not offer much differentiating power
\cite[e.g.,][]{Horne_etal_2004}. However, the different behaviors in
photocenter reverberation $x_0(t)$ for the flattened geometry and for the
bicone geometry might be useful in distinguishing the two geometries.

The astrometric reverberation mapping signals vanish when the iso-delay
surface is axisymmetric about the $z$-axis. For instance, an edge-on disk
will not produce photocenter offset as a function of time delay for the
velocity-integrated broad line flux. Fortunately, an astrometric signal will
still arise if we have velocity-resolved reverberation mapping: the blue side
of the reverberating broad line has a photocenter shifted from that of the
red side. Thus the astrometric feature can determine the angular momentum
direction of the disk-like BLR, while the traditional intensity reverberation
mapping is unable to derive such information.

Similarly, velocity-resolved reverberation mapping in flux amplitude and in
photocenter offset is a powerful tool to further constrain the velocity field
of the BLR. Below we use the three models to demonstrate the different
reverberation behaviors in the blueshifted and redshifted halves of the line
profile.

\begin{enumerate}

\item Spheroid models. Since we assumed a random Maxwellian velocity
    distribution, the red part and blue part of the emission line profile
    have identical behavior in reverberation signals.

\item Disk models. We consider the disk example of $i=45^\circ$ shown in
    Fig.\ \ref{fig:rm_examp}. In this case the redshifted part ($v_{\rm
    los}<0$) and blueshifted part ($v_{\rm los}>0$) of the disk will
    behave exactly the same in terms of $\delta_{\rm em}(t)$ and
    $x_0(t)$, except that the amplitude of $\delta_{\rm em}(t)$ is
    reduced by half. However, since we are now looking at half of the
    disk, the $y$ photocenter position of the redshifted/blueshifted disk
    changes as a function of time, as indicated by the red/blue lines in
    Fig.\ \ref{fig:rm_vel_examp}.

\item Bicone outflow models. We consider the bicone example of
    $i=45^\circ$ and $\Omega=\pi/2$ shown in Fig.\ \ref{fig:rm_examp}. In
    this example the near-side cone has blueshifted velocity while the
    far-side cone has redshifted velocity\footnote{If the half-opening
    angle of the bicone is larger than $90^\circ - i$ then the near/far
    side of the bicone will also contribute to redshifted/blueshifted
    velocities.}. We plot the results for both cones in Fig.\
    \ref{fig:rm_vel_bicone_examp}. The line intensity $\delta_{\rm em}$
    is already different for the blueshifted half and redshifted half,
    since the two cones have distinct los velocities: we see intensity
    reverberation from the blueshifted cone (near-side) earlier than the
    redshifted cone (far-side). The time evolution of the $x$ photocenter
    position is also different for the blueshifted and redshifted cones.
    The $y$ photocenter positions for the two cones are always zero due
    to the symmetry of this example.

\end{enumerate}

In the traditional case, velocity-resolved intensity RM can effectively
distinguish outflow/inflow kinematics from virial motion (either in a
Keplerian disk or in randomly orientated orbits) of the emission line region.
The distinction between the two kinds of kinematics is also clearly reflected
in the velocity-resolved astrometric RM, i.e., the time evolution of the
blueshifted and redshifted photocenters is asymmetric in the outflow/inflow
case, and symmetric in the case of virial motion (cf. Figs.\
\ref{fig:rm_vel_examp} and \ref{fig:rm_vel_bicone_examp})

\section{Discussion}\label{sec:disc}

\subsection{Practical Concerns with this Method}\label{sec:disc_prac}

The representative examples shown in \S\ref{sec:result} indicate that the
expected astrometric signals are on the order of the BLR size. Then for
typical quasar luminosities of $L_{\rm bol}\gtrsim 10^{46}\,{\rm
erg\,s^{-1}}$, we expect such astrometric signals to be on the order of tens
of $\mu$as at $z\sim 0.5$. For more luminous objects and/or at lower
redshifts, the expected astrometric signals will be larger. For example, the
quasar 3C 273 ($z=0.158, \lambda L_{\lambda}(5100\textrm{\AA})=8.6\times
10^{45}\,{\rm erg\,s^{-1}}, V\sim 12.8$) has a measured \hbeta\ BLR size of
$\sim 300\,$light days \citep[][]{Bentz_etal_2009a}, corresponding to $\sim
90\,\mu$as at its redshift. If we further consider reverberation of the dust
torus, then the expected astrometric signals would be even larger, e.g.,
$\gtrsim 100\,\mu$as for quasar luminosities. The sizes of the BLR and torus
approximately scale with $L^{0.5}$, thus for Seyfert luminosities, $\mu$as
astrometric precision might be necessary for astrometric RM.

However, even if the astrometry requirement is fulfilled (see
\S\ref{sec:disc_astro}), there are still several practical issues to
consider. These will affect the target selection and observing strategy for a
successful monitoring program. Detailed characterization of the adverse
effects and methods to mitigate them are beyond the scope of this paper, and
we only give a brief discussion here as a general guideline for future work.

First, the monitoring should be carried out in optical/near-IR to observe the
restframe UV to near-IR broad emission lines (e.g., \CIV, \MgII, Balmer and
Paschen lines) and adjacent continua. Tunable narrow-band filters are desired
for the broad line imaging and reliable continuum subtraction. High
photometric accuracy is generally required, and to detect and model the
astrometric signals we need good cadence of monitoring data. The monitoring
time baseline could span weeks to years, depending on the target.
Observations of repeated reverberation events can be combined to provide
improved model constraints, as long as the monitoring time span is much
shorter than the dynamical timescale ($\tau_{\rm dyn}\sim R/V_{\rm FWHM}$) of
the BLR.

Measuring the photocenter of the reverberating part of the BLR to the nominal
astrometric accuracy poses a more difficult challenge. It requires the
subtraction of the continuum and the steady-state BLR emission and subsequent
measurement of the photocenter of the variable BLR emission without degrading
the astrometric accuracy much. The use of narrow-band filters helps with this
procedure, but also increases the integration time to achieve the required
photometric precision. Take the broad \hbeta\ line for example, the typical
restframe equivalent width is $\sim 70\,$\AA\
\citep[e.g.,][]{Shen_etal_2011}. Thus for a $\sim 100\,$\AA\ narrow-band
filter centered on the redshifted \hbeta\ line at $z\sim 0.5$, the enclosed
\hbeta\ line flux is $\sim 50\%$ of the total flux. The host galaxy
contribution in this bandpass is negligible for quasar luminosities $\lambda
L_{\lambda}(5100\textrm{\AA})>10^{45}\,{\rm erg\,s^{-1}}$, and reaches $\sim
20\%$ of the total continuum at $\lambda
L_{\lambda}(5100\textrm{\AA})=10^{44.5}\,{\rm erg\,s^{-1}}$
\citep[e.g.,][]{Shen_etal_2011}\footnote{These host contamination estimates
were based on SDSS spectra with 3\arcsec\ diameter fibers. For our purposes,
the extraction aperture would be much smaller, so host contamination is
expected to be significantly less.}. With SNR$>100$ for the PSF core, $>10\%$
fractional changes in the broad line flux can be detected. However, the
centering accuracy of the variable BLR emission degrades as $\sigma_{\rm
meas,var}\sim \sigma_{\rm meas}/f_{\rm var}$, where $\sigma_{\rm meas}$ is
the nominal centering accuracy for the target and $f_{\rm var}$ is the
fraction of the variable flux (from the reverberating BLR) to the total
flux\footnote{For bright targets, however, the astrometric accuracy will be
limited by systematic errors rather than by centering accuracy (see
\S\ref{sec:disc_astro}).}. Thus larger amplitudes in the continuum variations
(hence larger reverberation amplitude) are always favored over smaller
amplitudes, just as for traditional intensity RM. Optimal image subtraction
methods may be required to isolate the variable BLR emission within the
bandpass \citep[e.g.,][]{Alard_2000}.


Finally, the real situation may deviate from the idealized case assumed in
reverberation mapping, which will inevitably lead to complications. With
better and better RM data sets, however, it is possible to incorporate
additional physical ingredients, such as photoionization processes,
absorption and scattering, more complex geometry/kinematics, etc., in the
framework of combined intensity and astrometric RM.

\subsection{Precision Astrometry in the Optical/Near-IR}\label{sec:disc_astro}

We now discuss realistic timescales for achieving the astrometric precision
($\mu$as to tens of $\mu$as) required to detect the astrometric reverberation
mapping signals. Such single-epoch astrometric precision is already under
serious considerations for ground-based and space-based facilities.

\subsubsection{ground-based facilities}

The recent development of AO-assisted, high-angular resolution, near-infrared
observations with large, single-aperture telescopes has dramatically improved
the astrometry precision. Single-measurement precision of $\sim
200-300\,\mu$as has been routinely achieved on $8-10$m class telescopes
\citep[e.g.,][]{Lazorenko_etal_2009,Fritz_etal_2010}. $\lesssim 100\,\mu$as
single-epoch precision (stable over 2 months) has been demonstrated for the
Palomar 200 inch Hale telescope in 2-minute exposures
\citep[][]{Cameron_etal_2009}, for stars with $K_s\lesssim 13$. At this level
of precision, there are several important instrumental, atmospheric and
astrophysical effects that need to be taken into account; and in most cases,
optimized procedures or algorithms must be undertaken to minimize these
effects \citep[e.g.,][]{Cameron_etal_2009,Fritz_etal_2010,Trippe_etal_2010}.

In the study by \citet{Cameron_etal_2009}, it was shown that the dominant
systematic error is due to atmospheric differential tilt jitter. This is the
stochastic and achromatic fluctuation in the relative displacement of the
target and the reference star due to the different columns of atmospheric
turbulence traversed by the light from the two objects.
\citet{Cameron_etal_2009} demonstrated that the tilt jitter effect can be
largely mitigated by an optimal weighting scheme that uses a grid of
reference stars to measure the pair-wise distances from the target. This
optimal estimation technique can essentially reduce the error due to tilt
jitter, $\sigma_{\rm TJ}$, below the centering accuracy $\sigma_{\rm meas}$,
which is mostly determined by aperture size, photon noise, and properties of
the PSF. \citet{Cameron_etal_2009} also provided an empirical formula to
predict the astrometric performance of a single-conjugate AO system:
\begin{eqnarray}
\sigma_{\rm tot}^2 = & & \sigma_{\rm meas}^2 + \sigma_{\rm TJ}^2 =\left(\frac{1.4\,{\rm sec}}{t}\right)\bigg\{\left[2\,{\rm mas}
\left(\frac{2}{N}\right)^{0.3}\left(\frac{5\,{\rm m}}{D}\right)^2\right]^2\nonumber \\
& & + \left[2\,{\rm mas}\left(\frac{2}{N}\right)^{0.7}\left(\frac{5\,{\rm m}}{D}\right)^{7/6}\right]^2\bigg\}\ ,
\end{eqnarray}
where $t$ is the integration time, $D$ is the telescope aperture diameter and
$N$ is the number of reference stars. This formula suggests that for
20-minute exposures, the single-measurement precision can achieve $\sim
20\,\mu$as for $10$m-class telescopes, and $< 10\,\mu$as for $30$m-class
telescopes, with only a few reference stars in a
$25$\arcsec$\times$$25$\arcsec\ FOV. Note that with these large-aperture
telescopes and a reasonable amount of exposure time (e.g., 30-60\,min), the
magnitude limit achievable with the above astrometric precision can be as
faint as $K_s\sim 18$ by extrapolation
\citep[e.g.,][]{Cameron_etal_2009,Trippe_etal_2010}, relevant for
reverberation monitoring of AGN continuum and broad line fluxes.

The above estimation might be optimistic. The actual astrometric performance
will inevitably depend on the telescope and AO systems, as well as target
properties. \citet{Trippe_etal_2010} performed a detailed study on the
limiting factors on the astrometry precision for the Multi-adaptive optics
Imaging CAmera for Deep Observations (MICADO) proposed for the 42-m European
Extremely Large Telescope (E-ELT). The dominant error terms include:
instrumental geometric distortion $\sigma_{\rm dist}\sim 30\,\mu$as,
chromatic atmospheric differential refraction (CDR) $\sigma_{\rm CDR}\sim
20\,\mu$as, atmospheric differential tilt jitter $\sigma_{\rm TJ}\sim
10\,\mu$as (with few minutes of integration), anisoplanatism of the
multi-conjugate AO system $\sigma_{\rm aniso}\sim 8\,\mu$as, using galaxies
as astrometric references in high galactic-latitude fields $\sigma_{\rm
galaxies}\sim 20\,\mu$as, and calibration of the projected pixel scale
$\sigma_{\rm scale}\lesssim 10\,\mu$as. Combining these error terms,
\citet{Trippe_etal_2010} estimated a total error budget of $\sigma_{\rm
sys}\sim 40\,\mu$as. In practice, some of these error terms can be reduced.
For instance, $\sigma_{\rm CDR}$ can be effectively suppressed if narrow-band
filters are used or special algorithms and observing strategies are deployed
to correct for the CDR \citep[e.g.,][]{Cameron_etal_2009}. Note that for BLR
reverberation mapping, it is preferable to use narrow-band filters to cover
the emission lines and nearby continuum. $\sigma_{\rm galaxies}$ can also be
reduced for AGN/quasar fields where there are a suitable number of reference
stars. It is also possible to correct instrumental distortion by dedicated
calibration procedures to control $\sigma_{\rm dist}$ down to the level of
$\sim 10-30\,\mu$as \citep[e.g.,][]{Trippe_etal_2010}. Thus the range of
single-epoch astrometry precision for this particular instrument on E-ELT is
$\sim 20-40\,\mu$as, although the lower bound would be very challenging to
achieve. We expect similar performance for other 30m-class telescopes.

An alternative route to single-aperture precision astrometry is long-baseline
astrometry that operates in the very-narrow-angle regime (angular separation
$\lesssim 30$\arcsec), which uses interferometry in optical/near-IR to
achieve high relative astrometric precision
\citep[e.g.,][]{Shao_Colavita_1992}. For instance, the Very Large Telescope
Interferometer (VLTI) instrument is designed to deliver an astrometric
precision at the $\sim 30-40\,\mu$as level \citep{van_Belle_2008}. In the
short term, however, the astrometric precision is likely limited, by the
accuracy in determining the interferometer baseline, to $\sim 20-30\,\mu$as
\citep[e.g.,][]{van_Belle_2009}. Moreover, current implementation of optical
to near-IR interferometry usually requires bright target magnitude $K\lesssim
13$ \citep[e.g.,][]{van_Belle_2008}, hence is not yet suitable for AGN
monitoring except for the brightest objects (such as 3C 273).

To summarize, ground-based facilities are promising to deliver $\lesssim$
tens of $\mu$as single-epoch astrometry precision within a decade or two, and
in the most optimal cases, $\sim 10\,\mu$as precision. Single-aperture
astrometry with large telescopes and AO (currently operating in the near-IR,
with potential extension to the optical) is probably better suited for AGN
targets than long-baseline interferometry at this stage. This precision can
be achieved for relatively short integration ($\lesssim 1\,$hr) and faint
target limit $K_s\lesssim 18$ relevant for AGN reverberation mapping. But to
achieve this goal, dedicated calibration algorithms and observing strategies,
along with systematic control of the telescope and instrument system, will
all be necessary. This level of astrometry precision is suitable for BLR
reverberation mapping of nearby AGNs and bright quasars, and sufficient for
AGN dust reverberation mapping.

\subsubsection{space-based facilities}

The {\em Hubble Space Telescope (HST)} is currently the only space-based
facility that can perform astrometry at the $\lesssim 1\,$mas precision
\citep[e.g.,][]{Anderson_King_2000,Benedict_etal_2003}. The future {\em James
Webb Space Telescope (JWST)} can in principle deliver a similar precision
\citep[e.g.,][]{Diaz-Miller_2007}. Thus both {\em HST} and {\em JWST} are
less favorable for precision astrometry compared with ground-based facilities
discussed above.

{\em GAIA} \citep[][]{Perryman_etal_2001} is a space-based astrometry mission
scheduled to launch in 2013, which can deliver 150 (400)\,$\mu$as
single-measurement astrometric precision for $V<16$ (18) targets. This
precision is not adequate for astrometric reverberation mapping of the BLRs,
and barely enough for AGN dust torus reverberation.

Perhaps the best facility for astrometric AGN reverberation mapping is a
mission similar to the SIM/SIM Lite mission \citep[][]{Unwin_etal_2008},
which is capable of achieving $\sim 10\,\mu$as single-measurement precision
for $V<18$ targets with a space-based 6-m baseline optical
interferometer\footnote{http://sim.jpl.nasa.gov/index.cfm}. In fact probing
the $\lesssim\,$pc spatial extent of AGN/quasar jets and BLRs is one of the
science cases for this mission, using the same technique as
spectroastrometry. Astrometric reverberation mapping would be an ideal
application with similar facilities for pointed multi-visit observations, yet
with much more constraining power on the BLR (and torus) geometry and
kinematics than spectroastrometry. Although this mission was canceled for the
current decade, the technology for such a mission is already in place
\citep[e.g.,][]{Shao_2010}. In light of the exciting science that will be
enabled with $\mu$as astrometry (especially with the driving force for
exoplanet science), it is reasonable to anticipate similar missions to be
approved in the next decade.

%

\section{Conclusions}\label{sec:con}

In this paper we outlined a simple idea of astrometric reverberation mapping:
observing the astrometric ``wobble'' of the BLR photocenter in reverberation
events. We demonstrated its potential in constraining the geometry and
kinematic structure of the BLR with simple BLR models.

The virtue of astrometric RM is that it provides an independent set of time
series that can be combined with the intensity RM data to constrain the BLR
model. In addition, it is more sensitive to the inclination of a disk-like
geometry\footnote{Other methods to infer the inclination of a disk geometry
of the BLR include polarimetry \citep[e.g.,][]{Li_etal_2009} and
double-peaked broad line profile fitting
\citep[e.g.,][]{Eracleous_Halpern_1994}.} and inference of its angular
momentum direction, than intensity RM alone. Although the BLR is still
unresolved, the spatial information from precision astrometry for the
reverberating parts of the BLR greatly facilitates using reverberation
mapping to probe the BLR structure. Once we have a set of high quality
continuum and broad line flux monitoring time series with precision
photometry and astrometry, we can fit the data in the time domain using model
intensity light curves and photocenter evolution. General Markov Chain Monte
Carlo approaches with Bayesian inference
\citep[e.g.,][]{Pancoast_etal_2011,Brewer_etal_2011} can be applied to search
the parameter space of different model families and to quantify model
uncertainties.

While conceptually simple, this method is not yet ready for immediate
applications. We discussed realistic timescales for achieving the required
astrometric precision ($\mu$as to tens of $\mu$as), and concluded that
ground-based large-aperture telescopes with AO offer the best hope to apply
this method to luminous quasars and dust torus RM within a decade or two, and
a space-based interferometer (a SIM/SIM-Lite like mission) with $\mu$as
astrometric precision in the next decade or two may extend this method to the
Seyfert regime.

\acknowledgements It is a pleasure to thank the anonymous referee for
suggestions that led to improvement of the paper, and Brad Peterson, Chris
Kochanek and Luis Ho for useful comments on the draft. Special thanks to Shu
Jia for a stimulating discussion that inspired this study. I gratefully
acknowledge support from the Smithsonian Astrophysical Observatory through a
Clay Postdoctoral Fellowship.

\end{document}